# On the accurate evaluation of overlap integrals over Slater type orbitals using analytical and recurrence relations

I.I. Guseinov

*Department of Physics, Faculty of Arts and Sciences, Onsekiz Mart University, Çanakkale, Turkey*

B.A. Mamedov

*Department of Physics, Faculty of Arts and Sciences, Gaziosmanpaşa University, Tokat, Turkey*

**Abstract**

In this study, using the analytical and recurrence relations suggested by the authors in previous works, the new efficient and reliable program procedure for the overlap integrals over Slater type orbitals (STOs) is presented. The proposed procedure guarantees a highly accurate evaluation of the overlap integrals with arbitrary values of quantum numbers, screening constants and internuclear distances. It is demonstrated that the computational accuracy of the proposed procedure is not only dependent on the efficiency of formulas, as has been discussed previously, but also on a number of other factors including the used program language package and solvent properties. The numerical results obtained using the algorithm described in the present work are in a complete agreement with those obtained using the alternative evaluation procedure. We notice that the program works without any restrictions and in all range of integral parameters.

## I. Introduction

In the study of the electronic structure of molecules, one has to evaluate overlap integrals over STOs accurately and efficiently. These integrals arise not only in the Hartree-Fock-Roothaan equations for molecules, but are also central to the calculation of arbitrary multicenter integrals based on the series expansion formulas about a new center and one-range addition theorems for STOs [1] which necessitate to accurately calculate the overlap integrals especially for the large quantum numbers. It should be noted that the overlap integrals over STOs are also used in all of the semiempirical methods [2].

The aim of this report is to calculate the overlap integrals over STOs using the analytical approach containing well-known auxiliary functions $A_k$ and $B_k$ and the recurrence relations for the basic overlap integrals presented in our previous works [3] and [4, 5], respectively. These expressions are especially useful for computation of overlap integrals on the computer for high quantum numbers, internuclear distances and orbital exponents or vice versa.

In this work, the differences and similarities in organization of existing overlap integral programs are discussed, and a new strategy is developed. This method is computationally simple and numerically well behaved. On the basis of formulas obtained in papers [3-5] we constructed a program for computation of the overlap integrals over STOs using Mathematica 5.0 international mathematical software and Turbo Pascal language packages. The numerical results demonstrated that the computational accuracy of the established formulas is not only dependent on the efficiency of formulas, but also strongly dependent on the used program language packages. Excellent agreement with benchmark results and stability of the technique are demonstrated. Since the overlap integrals over STOs are of considerable importance in the evaluation of arbitrary multicenter integrals, it is hoped that the present work will prove useful in tackling more complicated molecular integrals appearing in the determination of various properties for molecules when the Hartree-Fock-Roothaan approximation is employed.

## 2. Definition

The two-center overlap integrals over STOs with respect to lined-up coordinate systems are defined as

$$S_{nl\lambda,n'l'\lambda}(p,t) = \int \chi^*_{nlm}(\zeta,\vec{r}_a)\chi_{n'l'm}(\zeta',\vec{r}_b)dV, \qquad (1)$$

where $0 \leq \lambda \leq l, m = \pm\lambda, p = \dfrac{R}{2}(\zeta+\zeta'), t = (\zeta-\zeta')/(\zeta+\zeta'), \vec{R} \equiv \vec{R}_{ab} = \vec{r}_a - \vec{r}_b$ and

$$\chi_{nlm}(\zeta,\vec{r}) = (2\zeta)^{n+\frac{1}{2}}\left[(2n)!\right]^{-\frac{1}{2}} r^{n-1}e^{-\zeta r}S_{lm}(\theta,\varphi). \qquad (2)$$

Here, $S_{lm}$ is the complex $(S_{lm} = Y_{lm})$ or real spherical harmonic. It should be noted that our definition of phases for complex spherical harmonics $Y^*_{lm} = Y_{l-m}$ differs from the Condon-Shortley phases [6] by the sign factor.

## 3. Analytical relations in terms of auxiliary functions

In Ref.[3], using the auxiliary function method for the overlap integrals have been established the following formula:

$$S_{nl\lambda,n'l'\lambda}(p,t) = N_{nn'}(t) \sum_{\alpha=-\lambda}^{l}{}^{(2)} \sum_{\beta=\lambda}^{l'}{}^{(2)} g^0_{\alpha\beta}(l\lambda,l'\lambda) \sum_{q=0}^{\alpha+\beta} F_q(\alpha+\lambda,\beta-\lambda) \\ \times \sum_{m=0}^{n+n'-\alpha-\beta} F_m(n-\alpha,n'-\beta)A^{n+n'+1}_{n+n'-\alpha-\beta-m+q}(p)B_{m+q}(pt), \qquad (3)$$

where $N_{nn'}(t)$, $F_m(N,N')$ and $A^k_n(p)$ are determined by

$$N_{nn'}(t) = \dfrac{[(1+t)]^{n+1/2}[(1-t)]^{n'+1/2}}{\sqrt{(2n)!(2n')!}} \qquad (4)$$

$$F_m(N,N') = \sum_{\sigma=\frac{1}{2}[(m-n)+|m-n|]}^{\min(m,N')} (-1)^\sigma F_{m-\sigma}(N) F_\sigma(N'), \tag{5}$$

$$A_n^k(p) = p^k A_n(p). \tag{6}$$

Here, $F_m(n) = n!/[m!(n-m)!]$ are the binomial coefficients and $k \geq n+1$. It should be noted that, Eq.(5) for the generalized binomial coefficients with different notation $D_m^{NN'}$ firstly has been presented by N. Rosen in Ref. [7]. The quantities $A_n(p)$ and $B_n(pt)$ occurring in Eqs. (3) and (6) are well known auxiliary functions [8] (see also Ref. [9]).

The quantities $g_{\alpha\beta}^0(l\lambda, l'\lambda)$ in Eq.(3) are the expansion coefficients for a product of two normalized Legendre functions in elliptic coordinates. The relationship for these coefficients in terms of factorials was given in [10]. In Ref.[11], these coefficients were expressed in terms of binomial coefficients.

### 4. Use of recurrence relations for basic overlap integrals

In Ref.[5], using the expansion formula for product of two spherical harmonics both with the same center [10], the overlap integrals, Eq.(1), were expressed through the basic overlap integrals:

$$S_{nl\lambda,n'l'\lambda}(p,t) = \sum_{l''=\lambda}^{l} \frac{[2p(1+t)]^l}{[2p(1-t)]^{l''}} \left[ \frac{(2l+1)(2l'')! F_{2n'}(2n'+2l'') F_{l''+\lambda}(l+\lambda) F_{l''-\lambda}(l-\lambda)}{(2l''+1)(2l)! F_{2n-2l}(2n)} \right]^{1/2}$$
$$\times \sum_L \sqrt{2L+1} C^L(l'\lambda, l''\lambda) S_{n-l00,n'+l''L0}(p,t), \tag{7}$$

where $C^L(l'\lambda, l''\lambda)$ are the Gaunt coefficients. With the aid of recurrence relations given in Ref [5], the basic overlap integrals $S_{n00,n'l'0}(p,t)$ appearing in (7) can be expressed in terms of the functions $S_{00}(p,t) \equiv S_{000,000}(p,t)$ and $S_{00}(p,0) \equiv S_{000,000}(p,0)$ for the calculation of which we can use the following analytical formulas:

$$S_{00}(p,t) = \frac{1}{t} \eta_{00}(p,t) \{ e^{-p(1-t)} - e^{-p(1+t)} \} \tag{8}$$

$$S_{00}(p,0) = e^{-p}. \tag{9}$$

### 5. Numerical results and discussion

On the basis of Eqs.(3) and (7), obtained in our papers [3-5], we constructed the programs which were performed in the Mathematica 5.0 international mathematical software and Turbo Pascal 7.0 language packages. The computational results of overlap integrals by the use of Turbo Pascal 7.0 language package program have been examined in our published papers [3-

5]. The Barnett's data [12] and results of our calculation using Mathematica 5.0 international mathematical software and Turbo Pascal 7.0 language packages for various values of parameters are represented in Table 1. Barnett's data are reproduced by using our scheme with Mathematica while we get different results using the same scheme with Turbo Pascal. Thus, in this paper we show that the discrepancies can be arisen in the case of different programming environments. We note that, the difference between the numerical results of Eqs.(3) and (7) arise only after forty fifth digits. It should be noted that for the comparison of the accuracy of computer results obtained from the formulas of overlap integrals, one should use the same program language packages.

It is well known from the expert of this field that the problems occur in the evaluation of overlap integrals are as follow: small internuclear distances and small orbital exponents, and high internuclear distances and high orbital exponents. The results of calculation in these cases are given in Table 2. As is clear from our tests that the recurrence and analytical formulas presented in this study are useful tool for exact evaluation of the overlap integrals with arbitrary values of quantum numbers, internuclear distances and orbital parameters. Thus, our program calculates the overlap integrals over STOs with arbitrary quantum numbers $(n,l,n',l',\lambda)$ and variables (p,t).

**Table 1.** Comparison with results of Barnett [12]

| $n$ | $l$ | $n'$ | $l'$ | $\lambda$ | $p$ | $t$ | Eqs.(3) and (7) in Turbo Pascal procedure | Eqs.(3) and (7) in Mathematica procedure | Ref.[12] in Mathematica procedure |
|---|---|---|---|---|---|---|---|---|---|
| 3 | 2 | 3 | 2 | 1 | 25 | 0.6 | -4.42287766988261E-04 | -4.4228776698826080 6795415E-04 | -4.42287 76698 82608 80679E-04 |
| 4 | 2 | 4 | 3 | 1 | 80 | 0.4 | 4.03505950326382E-17 | 4.0350595032638229 810896077E-17 | 4.03505 95032 63822 98108E-17 |
| 5 | 4 | 5 | 4 | 4 | 100 | 0.7 | 1.56200599153976E-14 | 1.5620060274578910 37452179E-14 | 1.56200 60274 57891 03745E-14 |
| 7 | 3 | 4 | 3 | 2 | 150 | 0.7 | -1.76861050697887E-18 | -1.7686105069226485 90808884E-18 | -1.76861 05069 22648 59080E-18 |
| 9 | 5 | 8 | 4 | 3 | 45 | 0.2 | -5.46510243022867E-08 | -5.4651024302270401 73824997E-08 | -5.46510 24302 27040 17382E-08 |
| 10 | 7 | 8 | 2 | 1 | 60 | 0.2 | -1.84189026173558E-10 | -1.8418902617319810 64243984E-10 | -1.84189 02617 31981 06424E-10 |
| 10 | 9 | 10 | 9 | 9 | 15 | 0.6 | 6.23122318196866E-04 | 6.2312231819112494 64756102E-04 | 6.23122 31819 11249 46475E-04 |
| 13 | 12 | 13 | 12 | 12 | 25 | 0.01 | 1.35310560392189E-04 | 1.3531057870247123 81861868E-04 | 1.35310 57870 24712 38186E-04 |
| 14 | 13 | 14 | 13 | 13 | 15 | 0.4 | 4.53551312156525E-03 | 4.5355128510679091 15523032E-03 | 4.53551 28510 67909 11552E-03 |
| 15 | 14 | 15 | 14 | 14 | 15 | 0 | 3.74722497038009E-02 | 3.7472249703818919 54306084E-02 | 3.74722 49703 81891 95430E-02 |
| 16 | 15 | 16 | 15 | 15 | 35 | 0 | 1.21686562253236E-06 | 1.2168652185901881 88569061E-06 | 1.21686 52185 90198 18856E-06 |
| 17 | 8 | 8 | 7 | 4 | 50 | 0.1 | -1.00640061354258E-06 | -1.0064006411718817 23467400E-06 | -1.00640 06411 71881 72346E-06 |
| 17 | 16 | 17 | 16 | 16 | 25 | -0.5 | 3.06769565185575E-05 | 3.0677032557901936 09380388E-05 | 3.06770 32557 90193 60938E-05 |
| 18 | 12 | 18 | 12 | 12 | 20 | -0.6 | 6.63931813651240E-05 | 6.6393181369665067 75132120E-05 | 6.63931 81369 66506 77513E-05 |
| 21 | 10 | 9 | 8 | 6 | 45 | 0 | 5.38980685350612E-05 | 5.3898068533814377 30172720E-05 | 5.38980 68533 8143 773017E-05 |
| 27 | 8 | 9 | 8 | 7 | 35 | -0.2 | -1.73300982799699E-04 | -1.7442380751969590 91936618E-04 | -1.74423 80751 96959 09193E-04 |
| 30 | 10 | 14 | 10 | 8 | 35 | 0 | 1.35074709592800E-02 | 1.3507470959324333 88756335E-02 | 1.35074 70959 32433 38875E-02 |
| 37 | 8 | 12 | 10 | 6 | 10 | -0.6 | 3.98219849004259E-14 | 3.9822800437709157 35962091E-14 | 3.98228 00437 70915 73596E-14 |
| 40 | 4 | 12 | 4 | 3 | 15 | 0.6 | 9.48379265599810E-02 | 9.4837922083225567 85384419E-02 | 9.48379 22083 22556 78538E-02 |
| 43 | 10 | 18 | 8 | 6 | 60 | -0.4 | -1.15907687123104E-04 | -1.1582565326717481 46605545E-04 | -1.158256 53267 1748 14660E-04 |
| 50 | 4 | 50 | 4 | 4 | 25 | 0.7 | 1.84395901037228E-12 | 1.8439587993243634 03100208E-12 | 1.84395 87993 24363 40310E-12 |

**Table 2.** The comparative values of the two-center overlap integrals over STOs in lined-up coordinate systems for small and high values of integral parameters

| n | l | n' | l' | λ | p | t | Eqs.(3) and (7) in Mathematica procedure | Eqs.(3) and (7) in Turbo Pascal procedure |
|---|---|---|---|---|---|---|---|---|
| 7 | 4 | 7 | 4 | 4 | 0.01 | 0.01 | 0.999247898270316041412006 | 0.999247898270316 |
| 7 | 4 | 7 | 4 | 4 | 0.1 | 0.001 | 0.999757766779732929393514 | 0.999757766779732 |
| 7 | 4 | 7 | 4 | 4 | 0.01 | 0.001 | 0.999990152397157813463358966 | 0.999990152397158 |
| 7 | 4 | 7 | 4 | 4 | 0.0 | 0.0 | 1.000000000000000000000000 | 1.00000000000000000 |
| 7 | 4 | 7 | 4 | 4 | 0.001 | 0.1 | 0.927393290379437884684943 | 0.927393290379438 |
| 8 | 7 | 8 | 7 | 7 | 1E-4 | 1E-4 | 0.999999914705885568431130229397 | 0.999999914705885 |
| 8 | 7 | 8 | 7 | 7 | 1E-6 | 1E-6 | 0.999999999991470588235326272549 | 0.999999999991471 |
| 8 | 7 | 8 | 7 | 7 | 1E-6 | -0.5 | 0.086700327670739389394273246 4838 | 0.0867003276707394 |
| 10 | 9 | 10 | 9 | 9 | 1E-8 | 0.6 | 9.22337203685477579394533784 86E-03 | 9.22337203685478E-03 |
| 10 | 9 | 10 | 9 | 9 | 0.0 | 0.0 | 1.000000000000000000000000 | 1.000000000000000 |
| 10 | 9 | 10 | 9 | 9 | 1E-8 | 1E-8 | 0.999999999999998947619047619 05 | 0.9999999999999999 |
| 10 | 9 | 10 | 9 | 9 | 1E-5 | -0.8 | 2.1936950640359052899451116 4E-05 | 2.19369506403591 |
| 12 | 10 | 12 | 10 | 10 | 1E-5 | 1E-5 | 0.999999998748172653525286140114 | 0.999999998748173 |
| 12 | 10 | 12 | 10 | 10 | 1E-6 | 1E-6 | 0.999999999987481726528112453 | 0.999999999987482 |
| 12 | 10 | 12 | 10 | 10 | 1E-6 | 0.1 | 0.881941811798895655010568341338 | 0.881941811798896 |
| 12 | 10 | 12 | 10 | 10 | 1E-4 | -0.6 | 3.777893185901025124800447871E-03 | 3.77789318590103E-03 |
| 7 | 6 | 7 | 6 | 6 | 50 | 0.1 | 1.460223378297466376711404E-14 | 1.46022337769784E-14 |
| 10 | 9 | 16 | 10 | 9 | 60 | 0.1 | - 4.91326865764212881432637 55E-13 | - 4.91327027112068E-13 |
| 10 | 9 | 16 | 10 | 9 | 60 | 0.01 | -1.81096789566647263861898 93E-13 | -1.81096834940493E-13 |
| 13 | 10 | 13 | 10 | 10 | 35 | 0.1 | 9.7634850856025559477364730 5E-07 | 9.76348559116148E-07 |
| 7 | 4 | 7 | 4 | 4 | 100 | 0.1 | 2.72292316027798424617358955E-31 | 2.72292315888289E-31 |
| 75 | 30 | 75 | 20 | 18 | 1E-6 | 0.0 | -8.19297549621687882025926 3E-78 | -8.19297549621688E-78 |